\documentclass[12pt]{article}

\def\bn{\bigskip\noindent}

\usepackage{color}
\usepackage[dvips]{graphicx}
\usepackage{amsmath,amssymb}

\begin{document}
\begin{titlepage}
\renewcommand{\thefootnote}{\fnsymbol{footnote}}
 \font\csc=cmcsc10 scaled\magstep1
 {\baselineskip=14pt
 \rightline{
 \vbox{\hbox{December, 2003}
      \hbox{hep-th/0312090}
      }}}
\bn
\begin{center}
\Large{ \bf 
Topological Strings, Instantons and 
\\Asymptotic Forms
of Gopakumar--Vafa Invariants
}
\vspace*{1.5cm}

\normalsize{Yukiko Konishi}

\vspace*{1cm}

\normalsize{\it 
Research Institute for Mathematical Sciences, \\
Kyoto University, \\
Kyoto 606-8502, Japan}\\
\normalsize{\texttt{konishi@kurims.kyoto-u.ac.jp}}

\vspace{3ex}

\end{center}
\vspace{1cm}
\begin{abstract}
We calculate the topological string amplitudes 
of Calabi--Yau toric threefolds 
corresponding to 
$4D$, ${\mathcal N}=2$ $SU(2)$ gauge theory
with $N_f=0,1,2,3,4$
fundamental hypermultiplets
by using the method of  
the geometric transition
and show that they reproduce  
Nekrasov's formulas
for instanton counting.
We also determine the asymptotic
forms of the Gopakumar--Vafa invariants
of the Calabi--Yau threefolds including 
those at higher genera
from instanton amplitudes of the gauge
theory.
\end{abstract}
\end{titlepage}

\section{Introduction}
\hspace*{2.5ex}
Recently, remarkable developments
occurred in the theory of the topological strings.
We can now compute the Gromov--Witten
invariants or Gopakumar--Vafa invariants
of a Calabi--Yau toric threefold at all genera
by using the Feynman-like rules
\cite{AgMaVa, AgKlMaVa}
which has been developed from 
the geometric transition and the Chern--Simons theory
\cite{GoVa1}.
Although this method is 
most powerful compared to other methods such as  
localization and
local B-model calculation,
we still cannot  obtain the exact form of the 
topological string amplitude in general cases
because we have to sum over several partitions. 
However, it is found that 
we can perform the summation for 
some special types of tree graphs
by using the identities on the skew-Schur functions
\cite{EgKa, HoIqVa, Zhou1}.
A  simplest example  is 
the resolved conifold ${\mathcal O}(-1)\oplus
{\mathcal O}(-1)\to {\bf P}^1$.

Another  interesting application is the geometric engineering
of the gauge theories \cite{IqKa1, IqKa2, EgKa, HoIqVa}.
In this article we study the cases with the gauge group
$SU(2)$ and with $N_f=0,1,2,3,4$
fundamental hypermultiplets.
It has been known that
the corresponding Calabi--Yau threefolds  are the 
canonical bundles of the Hirzebruch surfaces 
blown up at $N_f$-points.
We calculate 
the topological string amplitudes of 
these Calabi--Yau threefolds
and show that they
reproduce Nekrasov's formulas for instanton counting
\cite{Nek}
in a certain limit. 
The $SU(n+1)$ cases without hypermultiplets and
the $SU(2)$ case with one hypermultiplet have been
studied in \cite{IqKa1, IqKa2, EgKa}
and the calculations in this article are 
essentially the same.
We also determine the asymptotic forms of the Gopakumar--Vafa
invariants of these Calabi--Yau threefolds
from the relation between the topological string amplitudes
and Nekrasov's formula.
This result is the generalization 
of the genus zero results 
\cite{KaKlVa, KoNa} to higher genus cases.

The organization of the paper is as follows.
In section \ref{topamp},
we calculate the topological string amplitudes
of Calabi--Yau toric threefolds that 
correspond to $4D$, ${\mathcal N}=2$ $SU(2)$
gauge theories with $N_f=0,1,2,3,4$ fundamental 
hypermultiplets.
In section \ref{nekrasov}, we
show that the topological amplitude 
reproduces Nekrasov's formula.
In section \ref{asymp},
we derive the asymptotic form of the Gopakumar--Vafa
invariants of the Calabi--Yau toric threefolds.
Appendices contain  formulas 
and the calculation of the framing.

\newcommand{\vsp}{\vspace*{1cm}}
\newcommand{\w}[1]{l(#1)}
\newcommand{\len}[1]{d(#1)}
\newcommand{\conj}[1]{{#1}^{t}}
\newcommand{\R}{R}
\newcommand{\Q}{Q}

\newcommand{\myp}[3]{P_{#1,#2}(#3)}


\section{Topological String Amplitude}
\label{topamp}
\hspace*{2.5ex}
In this section, we calculate the topological string
amplitudes of Calabi--Yau toric threefolds that 
reproduce four-dimensional ${\mathcal N}=2$
supersymmetric gauge theories with 
gauge group $SU(2)$ and with $N_f=0,1,2,3,4$
fundamental hypermultiplets.

\begin{figure}[t]
\begin{center}
{\scriptsize
\input{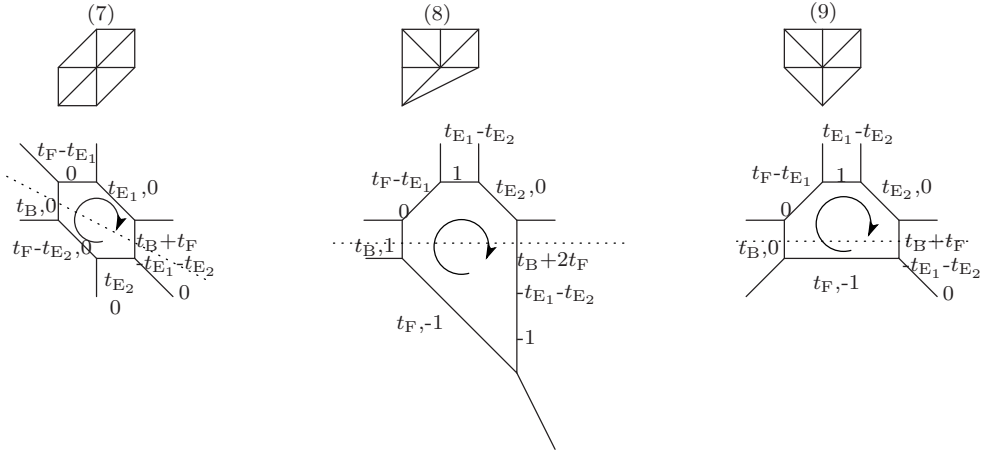}
}
\end{center}
\caption{The fans and the web diagrams
for the Calabi--Yau toric 3-folds that 
correspond to four-dimensional ${\mathcal N}=2$
$SU(2)$gauge theory with $N_f=2$ fundamental 
hypermultiplets.
For $N_f=0,1,2,4$ see figure \ref{loctv}.
}\label{nf2}
\end{figure}

First we briefly review the calculation of the 
topological string amplitudes of  
Calabi--Yau threefolds $X$ 
following  \cite{AgKlMaVa}
when $X$
is the canonical bundle of a smooth toric surface
classified in \cite{ChKlYaZa}.
Recall that 
a two-dimensional integral polytope 
(the section of the fan at the height 1) 
of $X$ has only one interior integral point  $(0,0)$
and this point and each integral point 
$v_i(1\leq i\leq k)$ on the boundary span an interior edge
(here we define $k$ to be the number of the interior edges and 
take $v_1,v_2,\ldots$ in the clockwise direction).
Therefore the corresponding web diagram consists of  
a polygon with $k$-edges
and external lines attached to it.
We take the orientation of edges on the polygon in
the clockwise direction, 
and that of the external lines 
in the outgoing direction. 
Then 
the integer $m_i$ of the framing for an interior edge 
dual to  $v_i$
is given by 
\begin{equation}\label{m}
m_i=-\gamma_i-1.
\end{equation}
Here $\gamma_i$ is 
the self-intersection number of the 
${\bf P}^1$ associated to $v_i$
and computed from the equation
\begin{equation}\label{SI}
-\gamma_i v_i=v_{i-1}+v_{i-1}.
\end{equation}
The derivation of $m_i$ is included in appendix.
Then we assign a  partition $\Q_i$ to each  interior edge
and the partition of zero to each external edge.
Finally we 
obtain the topological string amplitude 
by multiplying all quantities associated 
to vertices and edges and by
summing over all partitions. 
The brief summary of the rule is as follows: 
to a trivalent vertex
we associate
the three-point amplitude  
$C_{R_1,R_2,R_3}$ 
if the orientation of all the edges are outgoing,
$(-1)^{\w{\R_1}}C_{\conj{\R_1},\R_2,\R_3}$ if 
the orientation of one edge with a partition $\R_1$ 
is incoming, etc,
where
$\R_1,\R_2,\R_3$ are partitions 
assigned to three edges attached to the vertex;
to an interior edge we associate
$(-1)^{m\w{\R}}q^{-\frac{m\kappa(\R)}{2}}e^{-\w{\R}t}$
where $m$ is the integer coming from the framing
and $\R$ is the partition assigned to the edge and 
$t$ is the K\"ahler parameter of the corresponding ${\bf P}^1$.
Here $l(\R):=\sum_{i}\mu_i$ 
for a partition $\R=(\mu_1,\mu_2,\ldots)$ 
and $\kappa(\R):=\sum_{i}\mu_i(\mu_i-2i+1)$.
Thus the topological string amplitude for $X$ is written as
\begin{equation}
{\mathcal Z}=
\sum_{Q_1,\cdots,Q_k}\prod_{i=1}^k
C_{Q_i^t,\emptyset,Q_{i+1}}
(-1)^{\gamma_i \w{\Q_i}}
q^{\frac{\gamma_i+1}{2}\kappa(\Q_i)}.
\end{equation}
Here we have defined $\gamma_{k+1}:=\gamma_1,Q_{k+1}:=Q_1$.
This result was derived in \cite{Iqbal, Zhou, EgKa}.
The relation between the topological string amplitude
$\mathcal Z$
and the Gromov--Witten invariants is that 
$\log{\mathcal Z}|_{q=e^{\sqrt{-1}g_s}}$ is the
generating function of the Gromov--Witten invariants 
where $g_s$ is the genus expansion parameter.
This statement has been proved for
the canonical bundle of fano toric surfaces \cite{Zhou}
\footnote{${\bf P}^2$ and (2)(3)(5)(7) in figures \ref{nf2},
\ref{loctv}.}. 

Next we  derive more  compact formulas for the toric Calabi--Yau 
threefolds that 
corresponding to $4D$, ${\mathcal N}=2$ $SU(2)$
gauge theories with $N_f=0,1,2,3,4$ fundamental 
hypermultiplets. 
The Calabi--Yau threefolds are the canonical bundles of
the Hirzebruch surfaces ${\bf F}_0,{\bf F}_1$, or  ${\bf F}_2$
blown up at $N_f$ points.
There exist 3,2,3,3,2 such Calabi--Yau threefolds for 
$N_f=0,1,2,3,4$.\footnote{
Although there are 16 smooth toric  surfaces 
classified in \cite{ChKlYaZa},
three among them
($1$,$10,16$ in figure 1 in \cite{ChKlYaZa})
do not correspond to the four-dimensional gauge theories.}
The fans and the web diagrams for these threefolds are 
shown in figures \ref{nf2} and \ref{loctv}.
We take $[C_{\rm B}],[C_{\rm F}],[C_{{\rm E}_i}]$ 
$(1\leq i\leq N_f)$
as a basis of the second homology
where $C_{\rm B}$ (resp. $C_{\rm F}$)
is the base ${\bf P}^1$ (resp. fiber ${\bf P}^1$) 
and $C_{{\rm E}_i}$ is an exceptional curve. 
The intersections are 
\begin{equation}
\begin{split}
&C_{\rm B}.C_{\rm B}=-b,\quad
C_{\rm B}.C_{\rm F}=1,\quad
C_{\rm F}.C_{{\rm E}_i}=0 ,\\
&C_{\rm F}.C_{\rm F}=0,\quad
C_{\rm F}.C_{{\rm E}_i}=0 ,\quad
C_{{\rm E}_i}.C_{{\rm E}_j}=-\delta_{i,j}
\end{split}
\end{equation}
for $1\leq i,j\leq N_f$. 
The values of 
$b$ are $1$ or $2$ and will be listed later.
Here
$t_{\rm B},t_{\rm F},t_{\rm E_i}(1\leq i\leq 3)$
denote the K\"ahler parameters of the base ${\bf P}^1$,
 the fiber ${\bf P}^1$,  the $i$-th $(-1)$-curve
and 
$q_{\rm B}:=e^{-{t_{\rm B}}}$,
$q_{\rm F}:=e^{-{t_{\rm F}}}$, $q_i:=e^{-t_{\rm E_i}}$.

\newcommand{\B}{{$t_{\rm B}$}}
\newcommand{\F}{{$t_{\rm F}$}}
\newcommand{\Eu}{{$t_{\rm E_1}$}}
\newcommand{\Ed}{{$t_{\rm E_2}$}}
\newcommand{\Et}{{$t_{\rm E_3}$}}
\newcommand{\Eq}{{$t_{\rm E_4}$}}
\newcommand{\Ec}{{$t_{\rm E_5}$}}
\newcommand{\T}{{$t$}}

\begin{figure}[t]
\begin{center}
\input{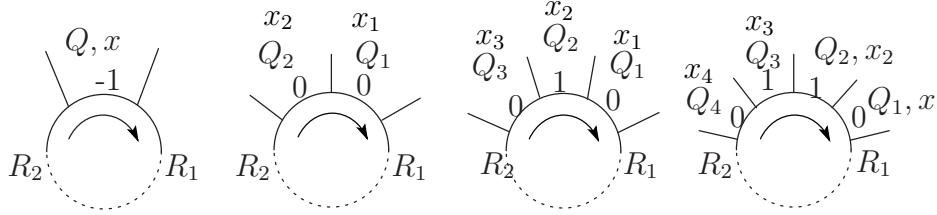}
\end{center}
\caption{The web diagrams corresponding to 
$H_{\R_1,\R_2}^{(k)}(x_1,\ldots,x_k)$
$(1\leq k\leq 4)$.}
\label{half}
\end{figure}

Now we compute the topological string amplitudes
by using the same strategy 
as \cite{IqKa1, IqKa2, EgKa}.
Let us take $N_f=2$ cases shown in figure
\ref{nf2} as examples.
We first cut the
polygon in the web diagram into two upper and 
lower parts (as shown by dotted line)
and compute the amplitudes separately.
Then we glue the two amplitudes together along the vertical
edges to obtain the whole topological string amplitude.
They are
written as follows:
\begin{align}
\mbox{(7):}&&
{\mathcal Z}&=\sum_{\R_1,\R_2}
H_{\R_1,\R_2}^{(2)}(q_1,q_{\rm F}{q_1}^{-1})
H_{\R_2,\R_1}^{(2)}(q_{\rm F}{q_2}^{-1},q_2)
\\\nonumber&&&\times
{q_{\rm B}}^{\w{\R_1}+\w{\R_2}}{q_{\rm F}}^{\w{\R_1}}
(q_1q_2)^{-\w{\R_1}},
\\
\mbox{(8):}&&
{\mathcal Z}&=\sum_{\R_1,\R_2}
H_{\R_1,\R_2}^{(3)}(q_2,q_1{q_2}^{-1},q_{\rm F}{q_1}^{-1})
H_{\R_2,\R_1}^{(1)}(q_{\rm F})
\\\nonumber&&&\times
(-1)^{\w{\R_1}+\w{\R_2}}
q^{\frac{\kappa(\R_1)}{2}-\frac{\kappa(\R_2)}{2}}
{q_{\rm B}}^{\w{\R_1}+\w{\R_2}}
{q_{\rm F}}^{2\w{\R_1}}
(q_1q_2)^{-\w{\R_1}},
\\
\mbox{(9):}&&
{\mathcal Z}&=\sum_{\R_1,\R_2}
H_{\R_1,\R_2}^{(3)}(q_2,q_1{q_2}^{-1},q_{\rm F}{q_1}^{-1})
H_{\R_2,\R_1}^{(1)}(q_{\rm F})
\\\nonumber&&&\times
{q_{\rm B}}^{\w{\R_1}+\w{\R_2}}
{q_{\rm F}}^{\w{\R_1}}
(q_1q_2)^{-\w{\R_1}}.
\end{align}
Here $\R_1$ (resp. $\R_2$) 
is a partition assigned to the right (resp. left)
vertical edge.
$H_{\R_1,\R_2}^{(k)}(x_1,\ldots,x_k)$ $(1\leq k\leq 4)$
are amplitudes corresponding to  
the web diagrams 
in figure \ref{half} and defined by
\begin{align}
H_{\R_1,\R_2}^{(1)}(x)&:=
\sum_{\Q}q^{\frac{\kappa(\Q)}{2}}x^{\w{\Q}}
(-1)^{\w{\R_2}}C_{\emptyset,\R_1,\conj{\Q}}
C_{\Q,\conj{\R_2},\emptyset},
\\
H_{\R_1,\R_2}^{(k)}(x_1,\ldots,x_k)&:=
\sum_{\Q_1,\cdots,\Q_k}(-1)^{\w{\R_2}+\w{\Q_1}+\w{\Q_k}}
q^{-\frac{\kappa(\Q_2)+\cdots+\kappa(\Q_{k-1})}{2}}
\prod_{i=1}^k {x_i}^{\w{\Q_i}} 
\\\nonumber 
&\times
C_{\emptyset,\R_1,\conj{\Q_1}}
C_{\conj{\Q_2},\emptyset,\Q_1}
\cdots
C_{\conj{\Q_{k}},\emptyset,\Q_{k-1}}
C_{\Q_k,\conj{\R_2},\emptyset}
\qquad(2\leq k\leq 4).
\end{align}
Using the expression for $C_{\R_1,\R_2,\R_3}$ written 
in terms of the skew-Schur functions and the identities
(see appendix), 
$H_{\R_1,\R_2}^{(k)}(x_1,\ldots,x_k)$
becomes 
\begin{equation}
\label{prop1}
\begin{split}
H_{\R_1,\R_2}^{(k)}(x_1,\ldots,x_k)&=
(-1)^{\w{\R_2}}W_{\R_1}W_{\conj{\R_2}}
\;K(x_1\cdots x_k)
g_{\R_1,\R_2}(x_1\cdots x_k)
\\&\times
\Big[\prod_{j=1}^{k-1}
K(x_1\cdots x_j)
g_{\R_1,\emptyset}(x_1\cdots x_j)
\prod_{j=2}^k
K(x_j\cdots x_k)
g_{\R_2,\emptyset}(x_j\cdots x_k)\Big]^{-1}
\\&\times
\prod_{2\leq i\leq j\leq k-1}
K(x_i\cdots x_j).
\end{split}
\end{equation}
The second (resp. third )
line  should be set to 1 for $k=1$ 
(resp. for $k\leq 2$).
\begin{align}
K(x)&:
=\sum_{i,j=1}^{\infty}(1-x^{-i-j+1})^{-1}
=\exp\Big[\sum_{k=1}^{\infty}\frac{q^kx^k}{k(q^k-1)}\Big]
\\
W_{\R}&:=q^{\frac{\kappa{\R}}{4}}
\prod_{1\leq i<j\leq \len{\R}}
\frac{[\mu_i-\mu_j+j-i]}{[j-i]}
\prod_{i=1}^{\len{\R}}
\prod_{v=1}^{\mu_i}
\frac{1}{[v-u+\len{\R}]},
\\\label{defg}
g_{\R_1,\R_2}(x)&:=
\prod_{i,j=1}^{\infty}\Big[
\frac{(1-x q^{\mu_{1,i}-i+\mu_{2,j}-j+1})}
{(1-x q^{-i-j+1})}\Big]
\\\label{numfree}&=\prod_{(i,j)\in\R_1}\frac{1}
{(1-x q^{\mu_{1,i}-j+\mu_{2,j}-i+1})}
\prod_{(j,i)\in\R_2}\frac{1}
{(1-x q^{-\mu_{1,j}^{\vee}+i-\mu_{2,i}^{\vee}+j-1})}.
\end{align}
Here $[k]:=q^{\frac{k}{2}}-q^{-\frac{k}{2}}$,
$l(\R):=\sum_{i}\mu_i$
for a partition $\R=(\mu_1,\mu_2,\ldots)$, 
$\kappa(\R):=\sum_{i}\mu_i(\mu_i-2i+1)$
and $d(\R)$ is the length of $\R$.
$(\mu^{\vee}_{1,i})_{i\geq 1}$
(resp. $(\mu^{\vee}_{2,i})_{i\geq 1}$)
is the conjugate partition of $\R_1$ (resp. $\R_2$)
and $(i,j)\in \R$ means that there is a box
in the place of $i$-th row and $j$-th column in 
the partition $\R$ regarded as a Young diagram. 
We have used an  identity in appendix 
to obtain the expression (\ref{numfree}). 
The final form of the topological string amplitudes
for the $N_f=2$ cases
become  
\begin{equation}
\begin{split}\label{top}
{\mathcal Z}&=Z_0\;Z_{\geq 1}\\
Z_0&=K(q_{\rm F})^2
\prod_{j=1}^{N_f}K(q_j)^{-1}K(q_{\rm F}{q_j}^{-1})^{-1}
\times
\begin{cases}
1&\mbox{(7)}\\
K(q_1{q_2}^{-1})&\mbox{(8)(9)}
\end{cases}
\\
Z_{\geq 1}&=
\sum_{\R_1,\R_2}
g_{\R_1,\conj{\R_1}}(1)g_{\R_2,\conj{\R_2}}(1)
g_{{\R_1},{\R_2}}(q_{\rm F})^2
\prod_{j=1}^{N_f}
g_{{\R_1},\emptyset}(q_j)^{-1}
g_{{\R_2},\emptyset}(q_{\rm F}{q_j}^{-1})^{-1}
\\
&\times {q_{\rm B}}^{\w{\R_1}+\w{\R_2}}
{q_{\rm F}}^{b \w{\R_1}}
(q_1\cdots q_{N_f})^{-\w{\R_1}}
(-1)^{m_1\w{\R_1}+m_2\w{\R_2}}
q^{-\frac{m_1\kappa(\R_1)+m_2\kappa(\R_2)}{2}}.
\end{split}
\end{equation}
The numbers $b$ are $1,2,1$ 
and $(m_1,m_2)$ are $(0,0),(-1,1),(0,0)$ 
for (7)(8)(9).
In $Z_{\geq 1}$ 
we have used the identity
\begin{equation*}
W_{\R}W_{\conj{R}}=(-1)^{\w{\R}}g_{\R_1,\conj{\R_1}}(1).
\end{equation*}

The generating function of the Gromov--Witten invariants
is obtained from the topological string amplitude
by taking the logarithm and substituting $e^{\sqrt{-1}g_s}$
into $q$:
\begin{equation}
\begin{split}
\log {\mathcal Z}|_{q=e^{\sqrt{-1}g_s}}&=
\sum_{g=0}^{\infty}{g_s}^{2g-2}
\sum_{d_{\rm B},d_{\rm F},d_1,d_2}
N_{g,d_{\rm B},d_{\rm F},d_1,d_2}
{q_{\rm B}}^{d_{\rm B}}
{q_{\rm F}}^{d_{\rm F}}
{q_1}^{d_{1}}
{q_{2}}^{d_{2}}
\\&=
\sum_{g=0}^{\infty}
\sum_{d_{\rm B},d_{\rm F},d_1,d_2}
\sum_{k=1}^{\infty}
\frac{n_{d_{\rm B},d_{\rm F},d_1,d_2}^g}{k}
\Big(2\sin\frac{k g_s}{2}\Big)^{2g-2}
\big({q_{\rm B}}^{d_{\rm B}}{q_1}^{d_1}
{q_{\rm F}}^{d_{\rm F}}
{q_1}^{d_{1}}{q_{2}}^{d_{2}}\big)^k.
\end{split}
\end{equation}
$N_{g,d_{\rm B},d_{\rm F},d_1,d_2}$ 
denotes the genus zero, 0-pointed Gromov--Witten invariant
for an integral homology class 
$d_{\rm B}[C_{\rm B}]+d_{\rm F}[C_{\rm F}]+
d_1[C_{E_1}]+d_2[C_{E_2}]$ and 
$n_{d_{\rm B},d_{\rm F},d_1,d_2}^g$
denotes the Gopakumar--Vafa invariant.

One can read off the Gromov--Witten invariants
with $d_{\rm B}=0$ from $Z_0$, because only
$\log Z_0$ gives the terms with degree zero in $q_{\rm B}$:
\begin{equation}
\begin{split}
\log Z_0&=
2\sum_{k=1}^{\infty}
\frac{q^k {q_{\rm F}}^k}{k(q^k-1)^2}
-\sum_{i=1}^{N_f}\Big[
\sum_{k=1}^{\infty}
\frac{q^k{q_i}^k}{k(q^k-1)^2}
+\sum_{k=1}^{\infty}
\frac{q^k{(q_{\rm F}{q_i}^{-1})}^k}{k(q^k-1)^2}
\Big]
\\
&+\sum_{k=1}^{\infty}\frac{q^k(q_1{q_2}^{-1})^k}{k(q^k-1)^2}
\mbox{ for (8)(9), } 0 \mbox{ for (7)}
\\
&\stackrel{q=e^{\sqrt{-1}g_s}}{=}
\sum_{k=1}^{\infty}
\frac{1}{k}
\Big(2\sin\frac{k g_s}{2}\Big)^{-2}
\Big[
-2 {q_{\rm F}}^k
+\sum_{i=1}^{N_f}\big(
{q_{\rm F}}^k+({q_{\rm F}}{q_i}^{-1})^k\big)
\Big]
\\
&+\sum_{k=1}^{\infty}
\frac{-1}{k}
\Big(2\sin\frac{k g_s}{2}\Big)^{-2}
\big({q_1{q_2}^{-1}}\big)^k
\mbox{ for (8)(9), } 0 \mbox{ for (7)}.
\end{split}
\end{equation}
Hence the nonzero Gopakumar--Vafa invariants
for a second homology class 
$d_{\rm B}[C_{\rm B}]+d_{\rm F}[C_{\rm F}]+
d_1[C_{E_1}]+d_2[C_{E_2}]$
with $d_{\rm B}=0$ are as follows
(we slightly change the notation and write the 
Gopakumar--Vafa invariant as $n_{\alpha}^g$ for a second homology 
class $\alpha$): 
\begin{equation}\label{rel}
n_{[C_{\rm F}]}^{g=0} =-2,\qquad
n_{[C_{{\rm E}_i}]}^{g=0}
=n_{[C_{\rm F}]-[C_{{\rm E}_i}]}^{g=0}=+1
\quad (1\leq i\leq N_f),\\
\end{equation}
and 
\begin{equation}\label{irrel}
n_{[C_{{\rm E}_1}]-[C_{{\rm E}_2}]}^{g=0}=-1 \mbox{ only for (8)(9)}. 
\end{equation}
The invariants  at $[C_{\rm F}],
[C_{{\rm E}_i}]$ and $[C_{\rm F}]-[C_{{\rm E}_i}]$ 
$(1\leq i\leq N_f)$ 
are the same in all of the three cases, while
the invariant at 
$[C_{{\rm E}_1}]-[C_{{\rm E}_2}]$ 
are different.
We will interpret  these results 
from the viewpoint of relation to the 
Seiberg--Witten prepotential of $SU(2)$ gauge theory
in the next section.

The Gopakumar--Vafa invariants with $d_{\rm B}\geq 1$
are given by $\log Z_{\geq 1}$.
We remark two properties.
The one is that the Gopakumar--Vafa invariant 
is nonzero only when 
\begin{equation}\label{reg1}
0\leq -d_i\leq d_{\rm B}\qquad (1\leq {}^{\forall}i\leq N_f).
\end{equation}
One could read this fact from the 
expression (\ref{top}) as follows.
The summand is the polynomial in ${q_1}^{-1},{q_2}^{-1}$
of degree at most $\w{\R_1}+\w{\R_2}$
given that $g_{\R,\emptyset}(x)^{-1}$ 
is the polynomial in $x$ of degree
$\w{\R}$.
Therefore the degree in ${q_i}^{-1}$ is always equal or smaller 
than the degree in $q_{\rm B}$,
and this fact persists  when we take the logarithm.
Thus the Gromov--Witten invariants 
are zero
unless the condition (\ref{reg1}) is satisfied,
and so are the Gopakumar--Vafa invariants.
The other property is that the Gopakumar--Vafa invariants are
symmetric with respect to $q_1,q_2$.
This follows from the invariance of 
$Z_{\geq 1}$ under the exchange of
$q_1$ and $q_2$.

Finally, we summarize the topological string amplitude 
for all the cases corresponding to 
$SU(2)$
gauge theory with $N_f=0,1,2,3,4$ hypermultiplets
shown in 
figures \ref{nf2}, \ref{loctv}.

\begin{figure}[h]
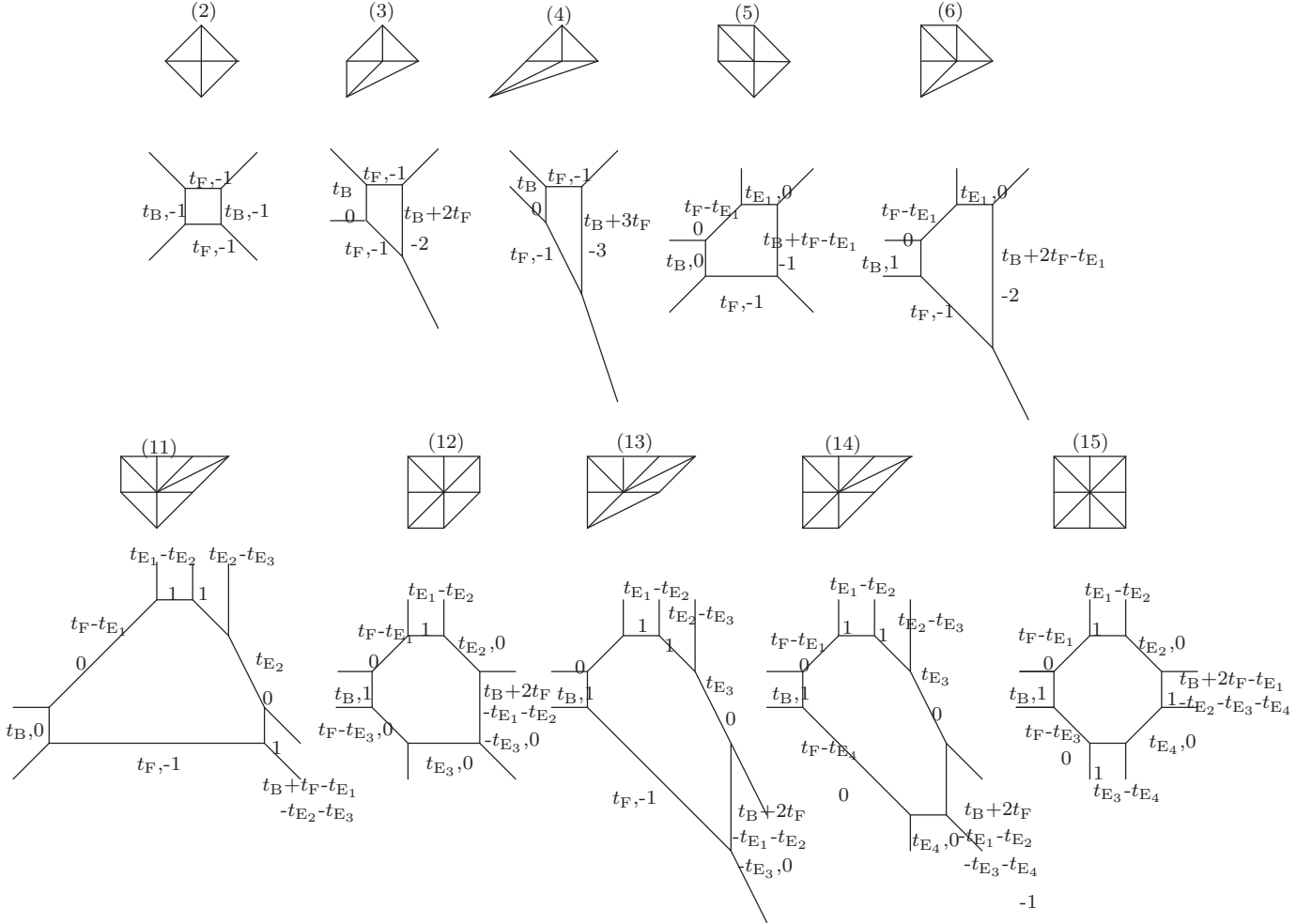

\begin{center}
{\scriptsize
\input{loctv.tex}
\input{loctv3.tex}
}\end{center}
\caption{The fans and the web diagrams
for the Calabi--Yau toric 3-folds that 
correspond to four-dimensional ${\mathcal N}=2$
$SU(2)$ gauge theory with $N_f$ fundamental 
hypermultiplets.
(2)(3)(4) are Hirzebruch surfaces and correspond to $N_f=0$,
(5)(6) to $N_f=1$, (7)(8)(9) to $N_f=2$
and (14)(15) to $N_f=4$.
The orientation of the interior edges are
taken in the clockwise direction.
$N_f=2$ cases are separately shown in figure \ref{nf2}.}
\label{loctv}
\end{figure}

\begin{equation}\label{gentop}
\begin{split}
{\mathcal Z}&=Z_0\;Z_{\geq 1}\\
Z_0&=K(q_{\rm F})^2
\prod_{j=1}^{N_f}K(q_j)^{-1}K(q_{\rm F}{q_j}^{-1})^{-1}
\times(a)
\\
Z_{\geq 1}&=
\sum_{\R_1,\R_2}
g_{\R_1,\conj{\R_1}}(1)g_{\R_2,\conj{\R_2}}(1)
g_{{\R_1},{\R_2}}(q_{\rm F})^2
\prod_{j=1}^{N_f}
g_{{\R_1},\emptyset}(q_j)^{-1}
g_{{\R_2},\emptyset}(q_{\rm F}{q_j}^{-1})^{-1}
\\
&\times {q_{\rm B}}^{\w{\R_1}+\w{\R_2}}
{q_{\rm F}}^{b \w{\R_1}}
(q_1\cdots q_{N_f})^{-\w{\R_1}}
(-1)^{m_1\w{\R_1}+m_2\w{\R_2}}
q^{-\frac{m_1\kappa(\R_1)+m_2\kappa(\R_2)}{2}}.
\end{split}
\end{equation}
where
\begin{equation*}
(a)=\begin{cases}
1 &(2)(3)(4)(5)(6)(7)\\
K(q_1{q_2}^{-1})&(8)(9)(12)\\
K(q_1{q_2}^{-1})K(q_1{q_3}^{-1})K(q_2{q_3}^{-1})
&(11)(13)(14)\\
K(q_1{q_2}^{-1})K(q_3{q_4}^{-1})&(15).
\end{cases}
\end{equation*}
$b$ is the self-intersection of the base ${\bf P}^1$
and $m_1(m_2)$ is the integer of the framing 
of the right (left) vertical edge:
\begin{equation*}
\begin{array}{|cc|rrr|rr|rrr|rrr|rr|}\hline
&&(2)&(3)&(4)&(5)&(6)&(7)&(8)&(9)&(11)&(12)&(13)&
(14)&(15)\\\hline
N_f&&\multicolumn{3}{|c|}{0}&
\multicolumn{2}{|c|}{1}&
\multicolumn{3}{|c|}{2}&
\multicolumn{3}{|c|}{3}&
\multicolumn{2}{|c|}{4}\\\hline
b&&0&1&2&1&2&1&2&1&2&2&1&1&1\\
m_1&&-1&-2&-3&-1&-2&0&-1&0&1&0&0&1&1\\
m_2&&-1&0&1&0&1&0&1&0&0&1&1&1&1\\\hline
\end{array}
\end{equation*}
\begin{table}[t]
\begin{equation*}
\begin{array}{|c|l|l|}\hline
N_f&       &\mbox{ Nonzero Gopakumar--Vafa invariants}\\\hline
0&(2)(3)(4)&n_{[C_{\rm F}]}^{g=0} \\\hline
1&(5)(6)&n_{[C_{\rm F}]}^{g=0} =-2,n_{[C_{{\rm E}_1}]}^{g=0}
=n_{[C_{\rm F}]-[C_{{\rm E}_1}]}^{g=0}=1
\\\hline
2&(7)&n_{[C_{\rm F}]}^{g=0} =-2,n_{[C_{{\rm E}_i}]}^{g=0}
=n_{[C_{\rm F}]-[C_{{\rm E}_i}]}^{g=0}=1
(1\leq i\leq 2)
\\\cline{2-3}
 &(8)(9)&n_{[C_{\rm F}]}^{g=0} =-2,n_{[C_{{\rm E}_i}]}^{g=0}
=n_{[C_{\rm F}]-[C_{{\rm E}_i}]}^{g=0}=1
(1\leq i\leq 2),
n_{[C_{{\rm E}_1}]-[C_{{\rm E}_2}]}=-1\\\hline
3&(11)(13)&n_{[C_{\rm F}]}^{g=0} =-2,n_{[C_{{\rm E}_i}]}^{g=0}
=n_{[C_{\rm F}]-[C_{{\rm E}_i}]}^{g=0}=1
(1\leq i\leq 3),
\\&&
n_{[C_{{\rm E}_1}]-[C_{{\rm E}_2}]}=
n_{[C_{{\rm E}_1}]-[C_{{\rm E}_3}]}=
n_{[C_{{\rm E}_2}]-[C_{{\rm E}_3}]}=
-1\\\cline{2-3}
&(12)&n_{[C_{\rm F}]}^{g=0} =-2,n_{[C_{{\rm E}_i}]}^{g=0}
=n_{[C_{\rm F}]-[C_{{\rm E}_i}]}^{g=0}=1
(1\leq i\leq 3),
n_{[C_{{\rm E}_1}]-[C_{{\rm E}_2}]}=-1\\\hline
4&(14)&n_{[C_{\rm F}]}^{g=0} =-2,n_{[C_{{\rm E}_i}]}^{g=0}
=n_{[C_{\rm F}]-[C_{{\rm E}_i}]}^{g=0}=1
(1\leq i\leq 4),
\\&&
n_{[C_{{\rm E}_1}]-[C_{{\rm E}_2}]}=
n_{[C_{{\rm E}_1}]-[C_{{\rm E}_3}]}=
n_{[C_{{\rm E}_2}]-[C_{{\rm E}_3}]}=
-1\\\cline{2-3}
&(15)&n_{[C_{\rm F}]}^{g=0} =-2,n_{[C_{{\rm E}_i}]}^{g=0}
=n_{[C_{\rm F}]-[C_{{\rm E}_i}]}^{g=0}=1
(1\leq i\leq 4),
n_{[C_{{\rm E}_1}]-[C_{{\rm E}_2}]}=
n_{[C_{{\rm E}_3}]-[C_{{\rm E}_4}]}=-1\\\hline
\end{array}
\end{equation*}
\caption{The nonzero Gopakumar--Vafa invariants 
for a second homology class 
$d_{\rm B}[C_{\rm B}]+d_{\rm F}[C_{\rm F}]+
d_1[C_{E_1}]+d_{N_f}[C_{E_{N_f}}]$
with $d_{\rm B}=0$.}
\label{Gopakuma}
\end{table}
Note that
the properties of the 
Gopakumar--Vafa invariants
mentioned in $N_f=2$ cases hold
in all $N_f$ cases:
the nonzero
Gopakumar--Vafa invariants
for a second homology class 
$d_{\rm B}[C_{\rm B}]+d_{\rm F}[C_{\rm F}]+
d_1[C_{E_1}]+d_{N_f}[C_{E_{N_f}}]$
with $d_{\rm B}=0$ are summarized in 
table \ref{Gopakuma}
and one can easily see that 
in all cases
$n_{[C_{\rm F}]}^0=-2$, 
$n_{[C_{{\rm E}_i}]}^0=n_{[C_{\rm F}]-[C_{{\rm E}_i}]}^0=1$
$(1\leq i\leq N_f)$;  
for $d_{\rm B}\geq 1$,
the Gopakumar--Vafa invariants is nonzero only if
$0\leq -d_i\leq d_{\rm B}(1\leq {}^{\forall} i\leq N_f)$ and
they are symmetric with respect to  $d_1,\ldots,d_{N_f}$.
Note also that 
$Z_{\geq 1}$'s of (7) and (9) (resp. (14) and (15))
are the same,
which means that the Gopakumar--Vafa  
invariants in cases (7) and (9) (resp. (14) and (15))
with $d_{\rm B}\geq 1$ are completely the same.

\section{Nekrasov's Formula}
\label{nekrasov}
\hspace*{2.5ex}
In this section we show that the topological string
amplitude ${\mathcal Z}$ 
gives  the one-loop corrections 
in the prepotential and Nekrasov's formula \cite{Nek}
for the $4D$, ${\mathcal N}=2$ $SU(2)$ gauge theory
with $N_f$ fundamental hypermultiplets
at a certain limit.
The argument in this section closely follows that
in \cite{IqKa1, IqKa2, EgKa}.

First let us identify the parameters in the 
two sides:
\begin{equation}
t_{\rm F}=-2 \beta a,\qquad
t_{{\rm E}_i}=-\beta (a+m_i)\quad (1\leq i\leq N_f),
\qquad
q=-\beta\hbar.
\end{equation}
Then the limit we should take is the limit $\beta\to 0$. 
Here $a=a_1=-a_2$ is the vacuum expectation value
of the complex scalar field in a gauge multiplets,
$m_i$ $(1\leq i\leq N_f)$'s are mass parameters of 
the fundamental hypermultiplets.
$g_s$ has been introduced as the genus expansion parameter.

Next,
let us show that the $t_{\rm B}$ independent part
$Z_0$  in the topological string amplitude
gives the perturbative one-loop correction
terms. 
We again take $N_f=2$ cases (7)(8)(9)  as examples.
Note that
\begin{equation*}
\lim_{\hbar=0}\frac{q^k}{(q^k-1)^2}|_{q=e^{-\beta\hbar}}=
\frac{1}{\beta^2\hbar^2}.
\end{equation*}
Then
\begin{equation}\label{oneloop}
\begin{split}
\lim_{\hbar=0}\hbar^2\log Z_{\geq 1}&=
2\sum_{k=1}^{\infty}
\frac{1}{k^3}\Big[e^{-2k\beta a}
-\sum_{i=1}^{N_f}\big(e^{-\beta(a+m_i)k}+e^{-\beta(a-m_i)k}
\big)
\Big]
\\
&+\sum_{k=1}^{\infty}\frac{1}{k^3}e^{-k\beta(m_1-m_2)}
\mbox{ for (8)(9), } 0 \mbox{ for (7)}.
\end{split}
\end{equation}
Each trilogarithm  corresponds to 
one logarithmic term in the Seiberg--Witten prepotential.
We can see this correspondence in the following way.
If we take the third derivative in $a$,
the right-hand side of (\ref{oneloop}) 
becomes
\begin{equation}
\begin{split}
&-\frac{8e^{-2\beta a}}{1-e^{-2\beta a}}+
\sum_{i=1}^{N_f}\Big(
\frac{e^{-\beta(a+m_i)}}{1-e^{-\beta(a+m_i)}}+
\frac{e^{-\beta(a-m_i)}}{1-e^{-\beta(a-m_i)}}\Big)
\\
&\stackrel{\beta\to0}{=}
-\frac{8}{a}+
\sum_{i=1}^{N_f}\Big(
\frac{1}{a+m_i}+
\frac{1}{a-m_i}\Big).
\end{split}
\end{equation}
In passing to the second line,
we have used the formula $\sum_{k=0}^{\infty}x^k=(1-x)^{-1}$.
This is exactly the third derivative of the
one-loop correction in the $SU(2)$ 
Seiberg--Witten prepotential
with $N_f$ fundamental hypermultiplets.
Note that the last term in (\ref{oneloop})
does not cause any problem
because such term depends only on mass
parameters, not on $a$.

Now we show that 
the logarithm of $t_{\rm B}$-dependent part $Z_{\geq 1}$ in 
(\ref{gentop}) gives
the instanton correction terms in the gauge theory.
More precisely we show that $Z_{\geq 1}$ becomes
Nekrasov's formula in the limit $\beta\to 0$.
We introduce the following function
for the sake of convenience: for two partitions 
$\R_1=(\mu_{1,i})_{i\geq 1}$ and
$\R_2=(\mu_{2,i})_{i\geq 1}$,
\begin{equation}
\begin{split}
P_{\R_1,\R_2}(x):=
&\prod_{(i,j)\in\R_1}\frac{1}
{\sinh \frac{\beta}{2}\big(
\tilde{a}+
\hbar({\mu_{1,i}-j+\mu_{2,j}^{\vee}-i+1})\big)}
\\&\times
\prod_{(i,j)\in\R_2}\frac{1}
{\sinh\frac{\beta}{2}\big(\tilde{a}+\hbar
({-\mu_{1,j}^{\vee}+i-\mu_{2,i}+j-1})\big)},
\end{split}
\end{equation}
where $\tilde{a},\hbar,\beta$ are 
defined by $q=e^{-\beta\hbar}$ and 
$x=e^{-\beta\tilde{a}}$. 
Then the following equations hold:
\begin{equation}
\begin{split}
P_{\R_1,\R_2}(x^{-1})&=
(-1)^{\w{\R_1}+\w{\R_2}}P_{\R_2,\R_1}(x),\\
P_{\conj{\R_1},\conj{\R_2}}(x,q)&=
P_{\R_2,\R_1}(x,q),\\
g_{\R_1,\R_2}(x)&=x^{-\frac{\w{\R_1}+\w{\R_2}}{2}}
2^{-\w{\R_1}-\w{\R_2}}q^{\frac{-\kappa(\R_1)-\kappa(\R_2)}{4}}
P_{\R_1,\conj{\R_2}}(x),
\\
P_{\R_1,\R_2}(x)&=
\prod_{k,l=1}^{\infty}
\frac{\sinh\frac{\beta}{2}\big(
\tilde{a}+\hbar(\mu_{1,k}-\mu_{2,l}+l-k)\big)}
{\sinh\frac{\beta}{2}\big(\tilde{a}+\hbar(l-k)\big)}.
\end{split}
\end{equation}
Therefore we can rewrite 
$Z_{\geq 1}$ as follows.
\begin{align}
\nonumber
Z_{\geq 1}&=
\sum_{\R_1,\R_2}
P_{\R_1,\R_2}({q_{\rm F}})
P_{\R_2,\R_1}({q_{\rm F}}^{-1})
P_{\R_1,\R_1}(1)P_{\R_2,\R_2}(1)
\\&\times
\prod_{j=1}^{N_f}
P_{\R_1,\emptyset}({q_j})
P_{\R_2,\emptyset}({q_{\rm F}}^{-1}{q_j})
\\&\times
2^{-(4-N_f)(\w{\R_1}+\w{\R_2})}
{q_{\rm B}}^{\w{\R_1}+\w{\R_2}}
{q_{\rm F}}^{b \w{\R_1}-2\w{\R_1}-2\w{\R_2}}
\prod_{j=1}^{N_f}{q_j}^{\frac{-\w{\R_1}+\w{\R_2}}{2}}
\nonumber\\&\times
(-1)^{m_1\w{\R_1}+m_2\w{\R_2}+N_f\w{\R_2}}
q^{-\frac{m_1\kappa(\R_1)-m_2\kappa(\R_2)}{2}
+\frac{(2-N_f)(-\kappa(\R_1)+\kappa(\R_2))}{4}}.
\nonumber\\
\label{expr}&=
\sum_{\R_1,\R_2}
\prod_{i,j=1,2}\prod_{k,l=1}^{\infty}
\frac{\sinh\frac{\beta}{2}\big(
a_{i,j}+\hbar(\mu_{i,k}-\mu_{j,l}+l-k)\big)}
{\sinh\frac{\beta}{2}\big(a_{i,j}+\hbar(l-k)\big)}
\\\nonumber&\times
\prod_{j=1}^{N_f}
\prod_{(k,l)\in\R_1}
\sinh\frac{\beta}{2}
\big(a_1+m_j+\hbar(k-l)\big)
\prod_{(k,l)\in\R_2}
\sinh\frac{\beta}{2}
\big(a_2+m_j+\hbar(k-l)\big)
\\\nonumber&\times
2^{-(4-N_f)(\w{\R_1}+\w{\R_2})}
{q_{\rm B}}^{\w{\R_1}+\w{\R_2}}
{q_{\rm F}}^{b \w{\R_1}-2\w{\R_1}-2\w{\R_2}}
\prod_{j=1}^{N_f}{q_j}^{\frac{-\w{\R_1}+\w{\R_2}}{2}}
\\\nonumber&\times
(-1)^{m_1\w{\R_1}+m_2\w{\R_2}+N_f\w{\R_2}}
q^{-\frac{m_1\kappa(\R_1)-m_2\kappa(\R_2)}{2}}
\end{align}
We have rewritten $\R_2$ in the summation as $\conj{\R_2}$  
and $a_{1,2}=a_1-a_2=2a=-a_{2,1}$.
Next we look into the limit $\beta\to0$.
The first line of (\ref{expr}) becomes
\begin{equation*}
\Big(\frac{\beta}{2}\Big)^{-4(\w{\R_1}+\w{\R_2})}
\prod_{i,j=1,2}\prod_{k,l=1}^{\infty}
\frac{
a_{i,j}+\hbar(\mu_{i,k}-\mu_{j,l}+l-k)}
{a_{i,j}+\hbar(l-k)}.
\end{equation*}
The second line becomes
\begin{equation*}
\hbar^{N_f(\w{\R_1}+\w{\R_2})}
\Big(\frac{\beta}{2}\Big)^{N_f(\w{\R_1}+\w{\R_2})}
\prod_{j=1}^{N_f}
\prod_{(k,l)\in\R_1}
\big(\frac{a_1+m_j}{\hbar}+(l-k)\big)
\prod_{(k,l)\in\R_2}
\big(\frac{a_2+m_j}{\hbar}+(l-k)\big).
\end{equation*}
The third line and the fourth become
\begin{equation*}
2^{-(4-N_f)(\w{\R_1}+\w{\R_2})}
{q_{\rm B}}^{\w{\R_1}+\w{\R_2}},\qquad
(-1)^{c(\w{\R_1}+\w{\R_2})}
\end{equation*}
where
\begin{equation}
c=\begin{cases}
1&
\mbox{for (2)(4)(5)(8)(9)(11)(14)(15)}\\
0&
\mbox{for (3)(6)(7)(12)(13)}.
\end{cases}
\end{equation}
Thus if we take the limit $\beta\to0 $ with
\begin{equation}\label{limit}
\begin{split}
&q_{\rm B}=(-1)^c \beta^{4-N_f}{\tilde{q}},
\\
&t_{\rm F}=-2 \beta a,\qquad
t_{{\rm E}_i}=-\beta (a+m_i)\quad (1\leq i\leq N_f),
\qquad
q=-\beta\hbar,
\end{split}
\end{equation}
the topological string amplitude (\ref{gentop})
becomes Nekrasov's formula for instanton 
counting \cite{Nek}:
\begin{equation}\label{agr}
\begin{split}
\lim_{\beta\to 0}\log Z_{\geq 1}
&={\mathcal Z}_{\rm Nekrasov}^{(N_f)}
\\&=\sum_{\R_1,\R_2}
(\tilde{q}\hbar^{N_f})^{\w{\R_1}+\w{\R_2}}
\prod_{i,j=1,2}\prod_{k,l=1}^{\infty}
\frac{
a_{i,j}+\hbar(\mu_{i,k}-\mu_{j,l}+l-k)}
{a_{i,j}+\hbar(l-k)}.
\\&\times
\prod_{i=1,2}\prod_{j=1}^{N_f}\prod_{k=1}^{\infty}
\frac{\Gamma(\frac{a_i+m_j}{\hbar}+\hbar(1+\mu_{i,k}-k))}
{\Gamma(\frac{a_i+m_j}{\hbar}+\hbar(1-k))}.
\end{split}
\end{equation}
Here the meaning of $\tilde{q}$ is that 
$\tilde{q}=\Lambda^{4-N_f}$ for 
$N_f=1,2,3$ and $\tilde{q}=e^{\sqrt{-1}\pi \tau}$ for 
$N_f=4$ where $\tau$ is the value of the moduli of the 
Seiberg--Witten curve when $m_1=\cdots=m_4=0$.

\section{Asymptotic Form of Gopakumar--Vafa Invariants}
\label{asymp}
\hspace*{2.5ex}
In this section we derive the asymptotic forms
of the Gopakumar--Vafa invariants of the 
Calabi--Yau toric threefolds that 
correspond to the $SU(2)$ gauge theory 
with $N_f=0,1,2,3,4$ hypermultiplets.
We derive first the asymptotic forms of
the Gromov--Witten invariants and 
then those of the Gopakumar--Vafa invariants.

Let us state the result:
for a second homology class $\alpha=
d_{\rm B}[C_{\rm B}]+d_{\rm F}[C_{\rm F}]+
d_1[C_{{\rm E}_1}]+\cdots+d_{N_f}[C_{{\rm E}_{N_f}}]
$ with $d_{\rm B}\geq 1$ and 
$0\leq -d_i\leq d_{\rm B}(1\leq {}^\forall i\leq N_f)$
\footnote{
The reason we deal with the Gopakumar--Vafa invariants 
at this range is 
that those for
$d_{\rm B}=0$ are exactly determined
by $Z_0$ (table \ref{Gopakuma})
and that
those at $d_{\rm B}\geq 1$
are zero unless 
$0\leq -d_i\leq d_{\rm B}
(1\leq {}^\forall i\leq N_f)$ 
(see previous section).}
,
\begin{equation}
n_{\alpha}^g
\sim
\frac{2^{4d_{\rm B}+2g-2}{\mathcal F}_{g,d_{\rm B}} 
}{(4d_{\rm B}+2g-3)!}
d_{\rm F}^{4d_{\rm B}+2g-3}
\prod_{i=1}^{N_f}
{}_{d_{\rm B}}C_{|d_i|}(-1)^{|d_i|},
\end{equation}
where 
${}_nC_k=\frac{n!}{k!(n-k)!}$ is the 
binomial coefficient.
${\mathcal F}_{g,k}$ is defined by 
\begin{equation}\label{instamp}
\log Z_{\rm Nekrasov}^{(0)}=
\sum_{k=1}^{\infty}\Lambda^{4k}\sum_{g=0}^{\infty}
(\sqrt{-1}\hbar)^{2g-2}
\frac{{\mathcal F}_{g,k}}{a^{4k+2g-2}}.
\end{equation}
This asymptotic form is valid in the region
\begin{equation}
d_{\rm F}\gg d_{\rm B}(g+1).
\end{equation}
Note that
the asymptotic form of the 
Gopakumar--Vafa invariants 
consists of two factors.
The one is the asymptotic form
of $N_f=0$ case which is a monomial 
in $d_{\rm F}$ with the prefactor
given by the 
instanton amplitude of the gauge theory.
This factor is common to 
all the $N_f$ cases
and the genus dependence appears 
only through this part.
The other factor is the binomial part
which represents 
the dependence
on $d_i $'s $(1\leq i\leq N_f)$.

For concreteness, we compute $\log {\mathcal Z}^{(0)}_{\rm Nekrasov}$
up to ${\mathcal O}(\Lambda^{16})$:
\begin{equation}
\begin{split}
&\log{\mathcal Z}^{(0)}_{\rm Nekrasov}
=
{\frac{2\Lambda^4}{{\hbar^2}{{{a_{12}}}^2}}}
\\&-
{\frac{{{\Lambda}^8}(2{\hbar^2}-5{{{a_{12}}}^2})}
{{\hbar^2}{{(\hbar-{a_{12}})}^2}{{{a_{12}}}^4}
{{(\hbar+{a_{12}})}^2}}}
\\&+
{\frac{16{{\Lambda}^{12}}(8{\hbar^4}-26{\hbar^2}{{{a_{12}}}^2}+
9{{{a_{12}}}^4})}{3{\hbar^2}{{(\hbar-{a_{12}})}^2}
{{(2\hbar-{a_{12}})}^2}{{{a_{12}}}^6}
{{(\hbar+{a_{12}})}^2}{{(2\hbar+{a_{12}})}^2}}}
\\&-
{\frac{{{\Lambda}^4}(10368{\hbar^{10}}
-59328{\hbar^8}{{{a_{12}}}^2}+
105356{\hbar^6}{{{a_{12}}}^4}-67461{\hbar^4}{{{a_{12}}}^6}+
17718{\hbar^2}{{{a_{12}}}^8}-1469{{{a_{12}}}^{10}})}{2{\hbar^2}
{{(\hbar-{a_{12}})}^4}{{(2\hbar-{a_{12}})}^2}
{{(3\hbar-{a_{12}})}^2}{{{a_{12}}}^8}
{{(\hbar+{a_{12}})}^4}{{(2\hbar+{a_{12}})}^2}
{{(3\hbar+{a_{12}})}^2}}}
\\&+{\mathcal O}(\Lambda^{20}).
\end{split}
\end{equation}
If one expand this further by $\hbar$,
one could obtain 
the coefficients
$2^{4k+2g-2}{\mathcal F}_{g,k}$
in table \ref{Fgk}.

\begin{table}[t]
\begin{equation*}
\begin{array}{|l|crrrrrrr|}\hline
&g&0&1&2&3&4&5&6\\\hline
k&& & & & & & &\\
1&&-2&0&0&0&0&0&0\\
2&&-5&8&-11&14&-17&20&-23\\
3&&-48&{\frac{1024}{3}}&-1872&9376&-{\frac{134608}{3}}&
   208704&-951440\\
4&&-{\frac{1469}{2}}&13176&-171201&1971646&
   -{\frac{42777927}{2}}&224106774&-2295588586\\
\hline
\end{array}
\end{equation*}
\caption{$2^{4k+2g-2}{\mathcal F}_{g,k}$
for small $g,k$.}
\label{Fgk}
\end{table}

In the rest of this section we explain the 
derivation of the asymptotic form.
In the previous section,
we showed that the 
topological string amplitudes
reproduce Nekrasov's partition functions
at the field theory limit (\ref{limit}).
By taking the logarithm of the equation (\ref{agr}),
the left-hand side is written as
\begin{equation}
\begin{split}\label{lhs}
&\sum_{d_{\rm B}= 1}^{\infty}
{q_{\rm B}}^{d_{\rm B}}
\sum_{g=0}^{\infty}{g_{s}}^{2g-2}
\sum_{d_{\rm F},d_1,\ldots,d_{N_f}}
N_{g,d_{\rm B},d_{\rm F},d_1,\ldots,d_{N_f}}
{q_{\rm F}}^{d_{\rm F}}
{q_{1}}^{\rm {d_1}}\cdots
{q_{N_F}}^{\rm {d_{N_f}}}
\end{split}
\end{equation}
On the other hand, the logarithm of the 
right-hand side 
takes the following form:
\begin{equation}\label{rhs}
\begin{split}
&\sum_{k=1}^{\infty}
{\tilde{q}}^k
\sum_{g=0}^{\infty}{(\sqrt{-1}\hbar)}^{2g-2}
\frac{B_{g,k}^{(N_f)}
\big(\frac{m_1}{a},\ldots,\frac{m_4}{a}\big)}
{a^{(4-N_f)k+2g-2}}.
\end{split}
\end{equation}
Here $B^{(N_f)}_{g,k}$ is  a polynomial in 
$N_f$ variables 
and it has 
the degree $k$ as a polynomial in each 
variable. Moreover,
the highest degree term is 
\begin{equation}\label{highest}
{\mathcal F}_{g,k}
\Big(\frac{m_1}{a}\cdots\frac{m_{N_f}}{a}\Big)^{k}
\end{equation}
where ${\mathcal F}_{g,k}$ is the same for 
all $N_f=0,1,2,3,4$. This is clear from the expression of
Nekrasov's partition function in eq. (\ref{agr}).
Therefore it 
is given by (\ref{instamp}). 
We can also understand this 
from the well-known fact that 
a gauge theory with fundamental hypermultiplets
reproduces the pure Yang-Mills theory
by decoupling hypermultiplets.

Let us adopt the following ansatz:
\begin{equation}\label{ansatz}
N_{g,d_{\rm B},d_{\rm F},d_1,\ldots,d_{N_f}}
\sim
r_{d_{\rm B}}^{(g)}(d_{\rm F})
\prod_{i=1}^{N_f}
{}_{d_{\rm B}}C_{|d_i|}(-1)^{|d_i|}
\qquad(0\leq -d_i\leq d_{\rm B}).
\end{equation}
By substituting the ansatz into (\ref{lhs})
and  identifying 
the parameters as (\ref{limit}) ,
we obtain
\begin{equation}\label{next}
\sum_{d_{\rm B}= 1}^{\infty}
\sum_{g=0}^{\infty}
{\tilde{q}}^{d_{\rm B}}
(\sqrt{-1}\hbar)^{2g-2}
\beta^{4d_{\rm B}+2g-2}
\sum_{d_{\rm F}}r_{d_{\rm B}}^{(g)}(d_{\rm F})
e^{-2\beta a d_{\rm F}}
\prod_{i=1}^{N_f}
(a+m_i)^{d_{\rm B}}.
\end{equation}
The last factors have appeared from 
\begin{equation*}
\prod_{i=1}^{N_f}\sum_{k_i=1}^{d_{\rm B}}
{}_{d_{\rm B}}C_{k_i}(-1)^{k_i}
{q_{i}}^{-{k_i}}
=\prod_{i=1}^{N_f}(1-{q_i}^{-1})^{d_{\rm B}}
\sim \prod_{i=1}^{N_f}
\beta^{d_{\rm B}} (a+m_i)^{d_{\rm B}}.
\end{equation*}
Then comparing (\ref{rhs})(\ref{highest}) and (\ref{next})
as  a series in $\hbar$ and $\Lambda$,
we obtain the condition which $r_{d_{\rm B}}^{(g)}
(d_{\rm F})$ 
must satisfy:
\begin{equation}
\begin{split} \label{cond}
\beta^{4k+2g-2}
\sum_{d_{\rm F}}r_{d_{\rm B}}^{(g)}(d_{\rm F})
e^{-t_{\rm F} d_{\rm F}}
&=\frac{{\mathcal F}_{g,k}}{{a}^{4k+2g-2}}
\\&
=(2\beta)^{4k+2g-2}
\frac{{\mathcal F}_{g,k}}{{t_{\rm F}}^{4k+2g-2}}.
\end{split}
\end{equation}
We have actually used only 
(\ref{highest}) in the comparison
and we will discuss this point shortly. 
In passing to the second line, 
we have  identified 
$a$ with $t_{\rm F}/2\beta$. 
It is clear that the powers of $\beta$ cancel out 
and this relation is 
independent of the limit $\beta\to 0$.
Then we replace the summation over $d_{\rm F}$
with the integration and regard the 
left-hand side of (\ref{cond}) as
the Laplace transform of
$r_{d_{\rm B}}^{(g)}(d_{\rm F})$
as a function in $d_{\rm F}$  to a function in 
$t_{\rm F}$.
Therefore by performing the inverse Laplace
transform on the right-hand side,
we obtain the following:
\begin{equation}
r_{d_{\rm B}}^{(g)}(d_{\rm F})\sim
\frac{2^{4d_{\rm B}+2g-2}}{(4d_{\rm B}+2g-3)!}
{\mathcal F}_{g,d_{\rm B}} 
d_{\rm F}^{4d_{\rm B}+2g-3}.
\end{equation}
This asymptotic form is valid only 
in the region
\begin{equation}\label{reg_gw}
d_{\rm F}\gg d_{\rm B}.
\end{equation}
In the derivation, 
we have used only the  term
with the highest degree in $m_1,\ldots,m_{N_f}$
in (\ref{rhs}).
As it turned out, this is suffice,
because the 
contributions
from the terms with lower degrees (in $m_i$)
are smaller: it would be  monomials in
$d_{\rm F}$ with 
degree smaller than $4d_{\rm B}+2g-3$. 
Thus such terms can be neglected 
since we consider the region where $d_{\rm F}$ 
is large.

Finally let us consider the asymptotic form
of the Gopakumar--Vafa invariants.
It is the same as the Gromov--Witten invariants,
because the contribution from the 
lower degree and lower genus Gromov--Witten
invariants can be neglected. However,
the region where the asymptotic form is valid
becomes more restricted. It is
\begin{equation}
d_{\rm F}\gg d_{\rm B}(g+1).
\end{equation}
Here we have added the factor $(g+1)$
in the right-hand side of (\ref{reg_gw})
because the number of lower degree/genus terms is $d_{\rm B}(g+1)$
and this number must be small enough compared to
$d_{\rm F}$.

\section{Example}

\begin{table}[t]
{\tiny
\begin{equation*}\!\!\!\!\!\!\!\!\!\!\!\!\!\!\!\!\!\!\!\!
\begin{array}{|l|rrrrrrrrrrrrrrrrrrrrrr|}\hline
&d_{\rm F}&0&1&2&3&4&5&6&7&8&9&10&
          11&12&13&14&15&16&17&18&19&20\\\hline
n_{1,d_{\rm F},0,0}^0&&
1&3&5&7&9&11&13&15&17&19&21&23&25&27&29&31&33&35&37&39&41
\\
n_{1,d_{\rm F},0,-1}^0&&
0&-2&-4&-6&-8&-10&-12&-14&-16&-18&-20&-22&-24&-26&
-28&-30&-32&-34&-36&-38&-40
\\
n_{1,d_{\rm F},-1,0}^0&&
0&-2&-4&-6&-8&-10&-12&-14&-16&-18&-20&-22&
-24&-26&-28&-30&-32&-34&-36&-38&-40
\\
n_{1,d_{\rm F},-1,-1}^0&&
0&1&3&5&7&9&11&13&15&17&19&21&23&25&27&29&31&33&35&37&39
\\\hline
\end{array}
\end{equation*}
}
{\tiny
\begin{equation*}\!\!\!\!\!\!\!\!\!\!\!\!\!\!\!\!\!\!\!\!
\begin{array}{|l|rrrrrrrrrrrrrrrrr|}\hline
&d_{\rm F}&0&1&2&3&4&5&6&7&8&9&10&
          11&12&13&14&15\\\hline
n_{2,d_{\rm F},0,0}^0&&
1&3&-1&-25&-101&-277&-631&-1265&-2323&-3981&-6469&-10057
&-15081&-21925&-31051&-42977\\
n_{2,d_{\rm F},0,-1}^0&&
0&-2&1&29&127&375&898&1876&3554&6252&10375&16423&25001&36829&
52752&73750\\
n_{2,d_{\rm F},0,-2}^0&&
0&0&0&-6&-32&-110&-288&-644&-1280&-2340&-4000
&-6490&-10080&-15106&-21952&-31080\\
n_{2,d_{\rm F},-1,0}^0&&
0&-2&1&29&127&375&898&1876&3554&6252&10375&16423&25001&36829&52752&
73750\\
n_{2,d_{\rm F},-1,-1}^0&&
0&1&-1&-31&-153&-491&-1249&-2731&-5361&-9703
&-16481&-26599&-41161&-61491&-89153&-125971\\
n_{2,d_{\rm F},-1,-2}^0&&
0&0&0&5&35&135&385&910&1890&3570&6270&10395&16445&25025&36855&
52780\\
n_{2,d_{\rm F},-2,0}^0&&
0&0&0&-6&-32&-110&-288&-644&-1280&-2340&-4000
&-6490&-10080&-15106&-21952&-31080\\
n_{2,d_{\rm F},-2,-1}^0&&
0&0&0&5&35&135&385&910&1890&3570&6270&10395&
16445&25025&36855&52780\\
n_{2,d_{\rm F},-2,-2}^0&&
0&0&0&0&-6&-32&-110&-288&-644&-1280&-2340&-4000&-6490&-10080&
-15106&-21952\\\hline
\end{array}
\end{equation*}}
{\tiny
\begin{equation*}\!\!\!\!\!\!\!\!\!\!\!\!\!\!\!\!\!\!\!\!
\begin{array}{|l|rrrrrrrrrrrrrrrrr|}\hline
&d_{\rm F}&0&1&2&3&4&5&6&7&8&9&10&
          11&12&13&14&15\\\hline
n_{2,d_{\rm F},0,0}^1&&
1&0&0&9&68&300&988&2698&6444&13916&27764&51963&92248&156648&256104&
405204\\
n_{2,d_{\rm F},0,-1}^1&&
0&0&0&-8&-72&-352&-1248&-3600&-8976&-20064
&-41184&-78936&-143000&-247104&-410176&-657696\\
n_{2,d_{\rm F},0,-2}^1&&
0&0&0&0&9&68&300&988&2698&6444&13916&27764&51963&92248&
156648&256104\\
n_{2,d_{\rm F},-1,0}^1&&
0&0&0&-8&-72&-352&-1248&-3600&-8976&-20064&-41184
&-78936&-143000&-247104&-410176&-657696\\
n_{2,d_{\rm F},-1,-1}^1&&
0&0&0&7&74&403&1544&4722&12324&28578&60456&118833&219934&
387101&652912&1061684\\
n_{2,d_{\rm F},-1,-2}^1&&
0&0&0&0&-8&-72&-352&-1248&-3600&-8976&-20064
&-41184&-78936&-143000&-247104&-410176\\
n_{2,d_{\rm F},-2,0}^1&&
0&0&0&0&9&68&300&988&2698&6444&13916&27764&51963&92248&156648&
256104\\
n_{2,d_{\rm F},-2,-1}^1&&
0&0&0&0&-8&-72&-352&-1248&-3600&-8976&-20064&-41184
&-78936&-143000&-247104&-410176\\
n_{2,d_{\rm F},-2,-2}^1&&
0&0&0&0&0&9&68&300&988&2698&6444&13916&27764&51963&92248&156648
\\\hline
\end{array}
\end{equation*}}
{\tiny
\begin{equation*}\!\!\!\!\!\!\!\!\!\!\!\!\!\!\!\!\!\!\!\!
\begin{array}{|l|rrrrrrrrrrrrrrrrr|}\hline
&d_{\rm F}&0&1&2&3&4&5&6&7&8&9&10&
          11&12&13&14&15\\\hline
n_{2,d_{\rm F},0,0}^2&&
0&0&0&0&-12&-116&-628&-2488&-8036&-22404&-55836&-127328
&-270088&-539416&-1023736&-1859632\\
n_{2,d_{\rm F},0,-1}^2&&
0&0&0&0&11&121&715&3025&10285&29887&77077&180895&393250&
802230&1550978&2863718\\
n_{2,d_{\rm F},0,-2}^2&&
0&0&0&0&0&-12&-116&-628&-2488&-8036&-22404&-55836
&-127328&-270088&-539416&-1023736\\
n_{2,d_{\rm F},-1,0}^2&&
0&0&0&0&11&121&715&3025&10285&29887&77077&180895&393250&802230&
1550978&2863718\\
n_{2,d_{\rm F},-1,-1}^2&&
0&0&0&0&-10&-124&-800&-3620&-12980&-39380&-105248&-254540
&-567710&-1184040&-2333760&-4382872
\\
n_{2,d_{\rm F},-1,-2}^2&&
0&0&0&0&0&11&121&715&3025&10285&29887&77077&180895&393250&
802230&1550978\\
n_{2,d_{\rm F},-2,0}^2&&
0&0&0&0&0&-12&-116&-628&-2488&-8036&-22404&-55836&-127328
&-270088&-539416&-1023736\\
n_{2,d_{\rm F},-2,-1}^2&&
0&0&0&0&0&11&121&715&3025&10285&29887&77077&180895&393250&802230&
1550978\\
n_{2,d_{\rm F},-2,-2}^2&&
0&0&0&0&0&
0&-12&-116&-628&-2488&-8036&-22404&-55836&-127328&-270088&-539416
\\\hline
\end{array}
\end{equation*}}
\caption{The Gopakumar--Vafa invariants of 
(7)(9) for $(d_{\rm B},g)=(1,0),(2,0),(2,1),(2,2)$.
Those with $(d_{\rm B},g)=(1,1),(1,2)$ are omitted because 
they are zero.
Note that the Gopakumar--Vafa invariants are 
symmetric with respect to $d_1,d_2$.}
\label{gvinv}
\end{table}
\begin{figure}[t]
\includegraphics[width=8cm]{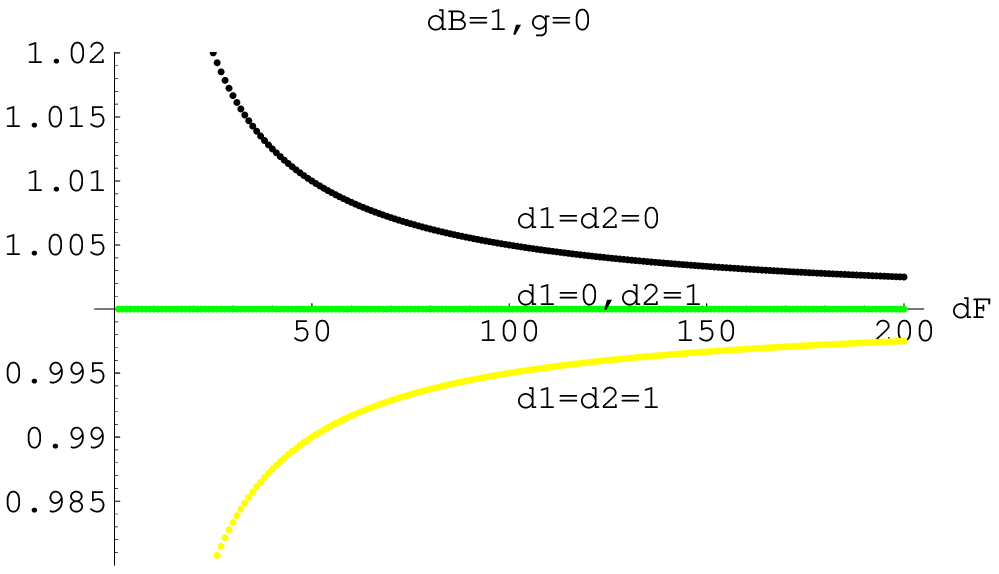}
\includegraphics[width=8cm]{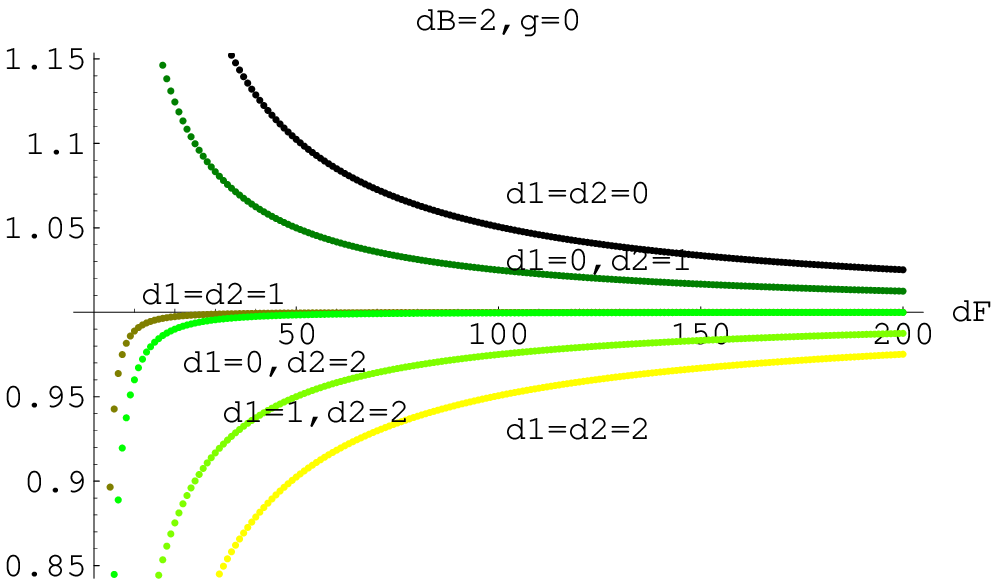}
\includegraphics[width=8cm]{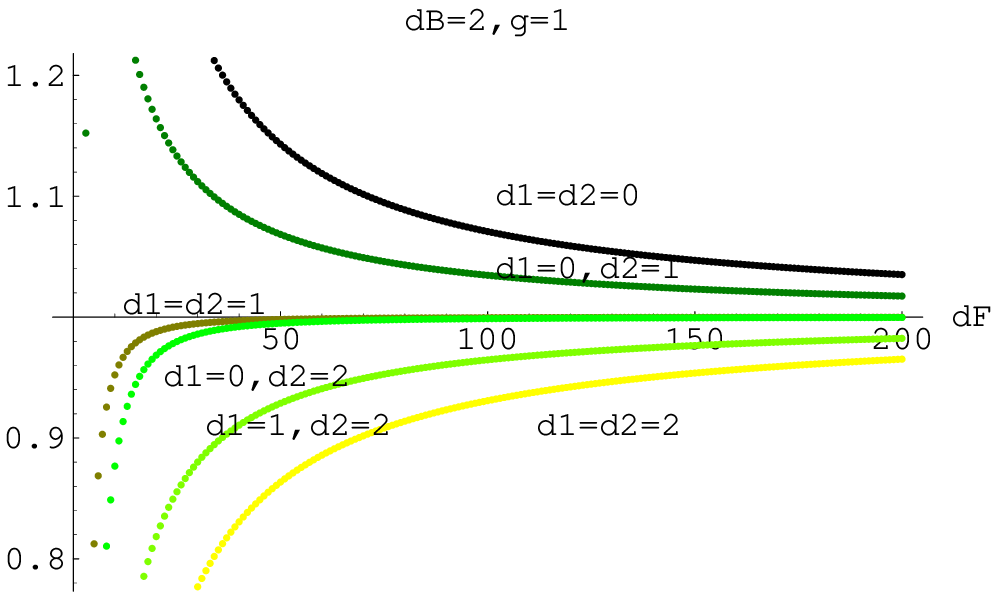}
\includegraphics[width=8cm]{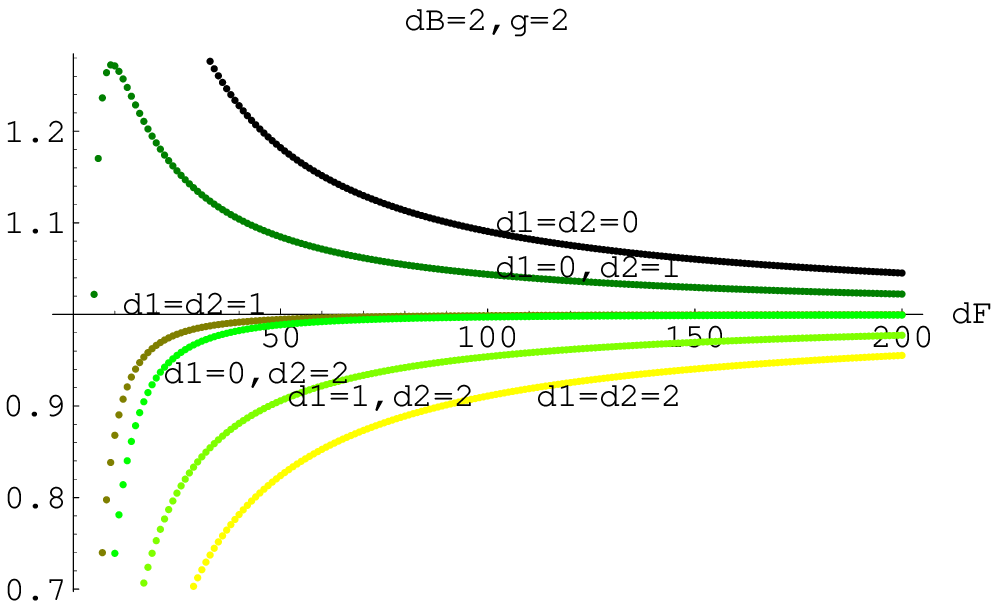}
\caption{The 
ratio between the Gopakumar--Vafa invariants 
$n_{d_{\rm B},d_{\rm F},d_1,d_2}^g$
and the asymptotic form in the case of (7)(9)
for $(d_{\rm B},g)=(1,0),(2,0),(2,1),(2,2)$.}
\label{ratio}
\end{figure}

\hspace*{2.5 ex}
In this section we explicitly compute
the Gromov--Witten invariants and compare
them with the asymptotic forms derived in the
previous section.
We take  as an example the case (7)(9) which correspond to
the $SU(2)$ gauge theory with $N_f=2$ 
fundamental hypermultiplets.
The ratios between the Gopakumar--Vafa invariants and 
the asymptotic forms are shown in figure \ref{ratio}
for $d_{\rm B}=1,2,g=0,1,2$.

We remark that the 
difference in the topological amplitudes 
of the two cases appears only in $Z_0$ and that
$Z_{\geq 1}$ are the same. 
Therefore the distribution of the 
Gopakumar--Vafa invariants $n_{d_{\rm B},d_{\rm F},d_1,d_2}^g$
for a homology class 
$\alpha=d_{\rm B}[C_{\rm B}]+d_{\rm F}[C_{\rm F}]+
d_1[C_{{\rm E}_1}]+d_2[C_{{\rm E}_2}]$ 
are the same when $d_{\rm B}\geq 1$.

Now we compute the Gopakumar--Vafa invariants 
of (7)(9) for $d_{\rm B}=1,2$, $g=0,1,2$.
Let $G_{d_{\rm B}}^{(g)}(d_{\rm F},d_1,d_2)$ denotes 
the generating function of the 
Gopakumar--Vafa invariants for given $d_{\rm B}$ and $g$:
\begin{equation}
G_{d_{\rm B}}^{(g)}(d_1,d_2):=\sum_{d_{\rm F},d_1,d_2}
n_{d_{\rm B},d_{\rm F},d_1,d_2}^g 
{q_{\rm F}}^{d_{\rm F}} 
{q_1}^{d_1}
{q_2}^{d_2}.
\end{equation} 
Here the summation over $d_{\rm F}$ is from zero to
infinity and 
the summations over $d_1,d_2$ are from $-d_{\rm B}$ to zero
(see section \ref{topamp}).
Then $G_{d_{\rm B}}^{(g)}$'s calculated from the 
topological string amplitude 
(\ref{top}) are as follows.
\begin{align*}
G_{1}^{(0)}&=
{\frac{{q_1}{q_2}+{q_F}-2{q_1}{q_F}-
2{q_2}{q_F}+{q_1}{q_2}{q_F}+{{{q_F}}^2}}
{{q_1}{q_2}{{(-1+{q_F})}^2}}},
\qquad
G_1^{(1)}=G_1^{(2)}=0,
\end{align*}
\begin{align*}
G_{2}^{(0)}&=\frac{\sum_{i,j=0}^2f_{ij}(q_{\rm F}){q_1}^i{q_2}^j}{
{{{q_1}}^2}{{{q_2}}^2}{{(-1+{q_F})}^6}
{{(1+{q_F})}^2} }
,\\
f_{00}&=-6{{{q_F}}^4}-8{{{q_F}}^5}-6{{{q_F}}^6},
\qquad
f_{01}=
5{{{q_F}}^3}+15{{{q_F}}^4}+15{{{q_F}}^5}+
5{{{q_F}}^6},
\\
f_{02}&=
-6{{{q_F}}^3}-8{{{q_F}}^4}-
6{{{q_F}}^5},
\qquad
f_{10}=
5{{{q_F}}^3}+15{{{q_F}}^4}+
15{{{q_F}}^5}+5{{{q_F}}^6},
\\
f_{11}&=
{q_F}-5{{{q_F}}^2}-23{{{q_F}}^3}-
29{{{q_F}}^4}-17{{{q_F}}^5}-7{{{q_F}}^6}-
{{{q_F}}^7}+{{{q_F}}^8},
\\
f_{12}&=
-2{q_F}+9{{{q_F}}^2}+17{{{q_F}}^3}+
7{{{q_F}}^4}+7{{{q_F}}^5}+4{{{q_F}}^6}-
2{{{q_F}}^7},
\\
f_{20}&=
-6{{{q_F}}^3}-8{{{q_F}}^4}-
6{{q_F}}^5,
\\
f_{21}&=
-2{q_F}+9{{{q_F}}^2}+
17{{{q_F}}^3}+7{{{q_F}}^4}+7{{{q_F}}^5}+
4{{{q_F}}^6}-2{{{q_F}}^7},
\\
f_{22}&=
1-{q_F}-9{{{q_F}}^2}-5{{{q_F}}^3}-
3{{{q_F}}^4}-3{{{q_F}}^5}-{{{q_F}}^6}+
{{{q_F}}^7}
\end{align*}
\begin{align*}
G_2^{(1)}&=\frac{\sum_{i,j=0}^2f_{ij}(q_{\rm F}){q_1}^i{q_2}^j}{
{{{q_1}}^2}{{{q_2}}^2}{{(-1+{q_F})}^8}
{{(1+{q_F})}^2}},
\\
f_{00}&=
9{{{q_F}}^5}+14{{{q_F}}^6}+9{{{q_F}}^7},
\qquad
f_{01}=
-8{{{q_F}}^4}-24{{{q_F}}^5}-24{{{q_F}}^6}-8{{{q_F}}^7},
\\
f_{02}&=
9{{{q_F}}^4}+14{{{q_F}}^5}+9{{{q_F}}^6},
\qquad
f_{10}=
-8{{{q_F}}^4}-
24{{{q_F}}^5}-24{{{q_F}}^6}-8{{{q_F}}^7},
\\
f_{11}&=
7{{{q_F}}^3}+32{{{q_F}}^4}+50{{{q_F}}^5}+
32{{{q_F}}^6}+7{{{q_F}}^7},
\qquad
f_{12}=
-8{{{q_F}}^3}-24{{{q_F}}^4}-24{{{q_F}}^5}-
8{{{q_F}}^6},
\\
f_{20}&=
9{{{q_F}}^4}+14{{{q_F}}^5}+9{{{q_F}}^6},
\qquad
f_{21}=
-8{{{q_F}}^3}-24{{{q_F}}^4}-24{{{q_F}}^5}-8{{{q_F}}^6},
\\
f_{22}&=
1-6{q_F}+13{{{q_F}}^2}+{{{q_F}}^3}+
37{{{q_F}}^5}-14{{{q_F}}^6}-8{{{q_F}}^7}+
13{{{q_F}}^8}-6{{{q_F}}^9}+{{{q_F}}^{10}}
\end{align*}
\begin{align*}
G_2^{(2)}&=-\frac{{q_{\rm F}}^4\sum_{i,j=0}^2f_{ij}(q_{\rm F}){q_1}^i{q_2}^j}{
{{{q_1}}^2}{{{q_2}}^2}{{(-1+{q_F})}^{10}}{{(1+{q_F})}^2}}
,\\
f_{00}&=12{{{q_F}}^2}+20{{{q_F}}^3}+12{{{q_F}}^4},
\qquad
f_{01}=-11{q_F}-33{{{q_F}}^2}-33{{{q_F}}^3}-
11{{{q_F}}^4},
\\
f_{02}&=
12{q_F}+20{{{q_F}}^2}+12{{{q_F}}^3},
\qquad
f_{10}=
-11{q_F}-33{{{q_F}}^2}-
33{{{q_F}}^3}-11{{{q_F}}^4},
\\
f_{11}&=
10+44{q_F}+68{{{q_F}}^2}+44{{{q_F}}^3}+
10{{{q_F}}^4},
\qquad
f_{12}=
-11-33{q_F}-33{{{q_F}}^2}-11{{{q_F}}^3},
\\
f_{20}&=
12{q_F}+20{{{q_F}}^2}+12{{{q_F}}^3},
\qquad
f_{21}=
-11-33{q_F}-33{{{q_F}}^2}-11{{{q_F}}^3},
\\
f_{22}&=
12+20{q_F}+12{{{q_F}}^2}.
\end{align*}
By expanding these as series in $q_{\rm F},{q_1}^{-1},{q_2}^{-1}$
we obtain the Gopakumar--Vafa invariants 
listed in table \ref{gvinv}.

\section{Conclusion}
\hspace*{2.5ex}
In this article, we have computed the topological string 
amplitudes of the canonical bundles of  toric surfaces
which are the Hirzebruch surfaces blown up at $N_f=0,1,2,3,4$
points and 
showed that in a certain limit, they reproduce the 
Nekrasov's formulas for
$4D$ ${\mathcal N}=2$ $SU(2)$
gauge theories with $N_f$ fundamental hypermultiplets.

We have  also derived the asymptotic form of the 
Gopakumar--Vafa invariants at all genera 
from the instanton amplitudes of 
the gauge theory.
From the result in \cite{KoSa} and ours,
we expect that the asymptotic form of 
the Gopakumar--Vafa invariants in  the $SU(n+1)$ cases   
are given as the product of 
the two factors: 
the one is
the asymptotic form of $N_f=0$ case
derived in \cite{KoSa}
and the other is just the  same binomial part 
as the $SU(2)$ case.

\section*{Acknowledgement}
\hspace*{2.5ex}
The author would like to thank
H. Kanno for valuable discussions.
She is also grateful 
to M. Naka and K. Sakai for 
collaboration in previous works. 

\appendix
\section{Formulas}
\hspace*{2.5ex}
We list some formulas in this section.

We use letters $\R,\R_i,R',Q$ etc. for partitions.
As mentioned before,
$l(\R):=\sum_{i}\mu_i$
for a partition $\R=(\mu_1,\mu_2,\ldots)$, 
$\kappa(\R):=\sum_{i}\mu_i(\mu_i-2i+1)$
and $d(\R)$ is the length of $\R$.
$(\mu^{\vee}_{1,i})_{i\geq 1}$
(resp. $(\mu^{\vee}_{2,i})_{i\geq 1}$)
is the conjugate partition of $\R_1$ (resp. $\R_2$)
and $(i,j)\in \R$ means that there is a box
in the place of $i$-th row and $j$-th column in 
the partition $\R$ regarded as a Young diagram.

The three point amplitude is \cite{AgKlMaVa, EgKa, HoIqVa, Zhou1}
\begin{equation}
\begin{split}
C_{\R_1,\R_2,\R_3}&:=
\sum_{Q_1,Q_3}N_{Q_1,Q_3^{t}}^{R_1,R_3^{t}}
q^{\kappa(R_2)/2+\kappa(R_3)/2}
\frac{W_{R_2^{t},Q_1}W_{R_2,Q_3^{t}}}
{W_{R_2,\emptyset}}
\\
&=(-1)^{\w{\R_2}}q^{\frac{\kappa{\R_3}}{2}}
s_{\conj{\R_2}}\sum_{Q}s_{\R_1/Q}(q^{\conj{\R_2}+\rho})
s_{\conj{\R_3}/Q}(q^{\conj{\R_2}+\rho}).
\end{split}
\end{equation}
In the first line the summation  
is over pairs of partitions $Q_1,Q_3$.
The coefficient $N_{Q_1,Q_2}^{R_1,R_3}$
is defined as follows:
\begin{equation}
N_{Q_1,Q_2}^{R_1,R_2}:=
\sum_{R}
N_{R,Q_1}^{R_1}N_{R,Q_2}^{R_2}.
\end{equation}
$N_{R,R''}^{R'''}$ is the tensor product coefficient.
In the second line, $s_{\R},s_{\R/Q}$ are
the Schur function and the skew Schur function \cite{Mac}:
\begin{align}
s_{\R}(x)&:=\frac{
\det({x_i}^{\mu_j+\len{\R}-j})_{1\leq i,j\leq \len{\R}}}
{\det({x_i}^{\len{\R}-j})_{1\leq i,j\leq \len{\R}}},
\\
s_{\R/Q}&:=\sum_{\R_1}N_{Q,\R_1}^{\R}s_{\R_1},
\end{align}
where $x=(x_1,\ldots,x_{\w{\R}})$.
The tensor product coefficient $N_{R,R''}^{R'''}$
can be computed from the formula 
\begin{equation}
s_{\R_1}s_{\R_2}=\sum_{Q}N_{\R_1,\R_2}^{Q}s_Q
\end{equation}
or by using the Littlewood-Richardson rule.

The  formulas  essential to the calculation of 
$H_{\R_1,\R_2}^{(k)}$ in section \ref{topamp} are
\cite{Mac}
\begin{align}
\sum_{Q}s_{Q/\R_1}(x)s_{\conj{Q}/\R_2}(y)&=
\prod_{i,j=1}^{\infty}
(1+x_iy_i)\sum_{Q}s_{\conj{R_2}/Q}(x)
s_{\conj{\R_1}/\conj{Q}},\\
\sum_{Q}s_{Q/\R_1}(x)s_{{Q}/\R_2}(y)&=
\prod_{i,j=1}^{\infty}
(1-x_iy_i)^{-1}\sum_{Q}s_{{R_2}/Q}(x)
s_{{\R_1}/{Q}}.
\end{align}
Note that the summations in the right-hand sides 
are infinite 
but those in the left-hand sides are finite.

Let $f(x)$ be a function in one variable.
For a partition $\R=(\mu_i)_{i\geq 1}$
($\conj{\R}=(\mu_i^{\vee})_{i\geq 1}$),
the following identities hold:
\begin{align}\label{start}
\prod_{1\leq i< j\leq \infty}
\frac{f(\mu_i-\mu_j+j-i)}{f(j-i)}
&
=\prod_{1\leq i<j\leq \len{\R}}
\frac{f(\mu_i-\mu_j+j-i)}{f(j-i)}
\prod_{i=1}^{\len{\R}}
\prod_{v=1}^{\mu_i}
\frac{1}{f(v-i+\len{\R})}
\\&
=\prod_{(k,l)\in\R}\frac{1}{f(\mu_k+\mu_l^{\vee}-k-l+1)}. 
\end{align}
For two partitions $\R_1=(\mu_{1,i})_{i\geq 1}$ and
$\R_2=(\mu_{2,i})_{i\geq 1}$,
the following identities hold:
\begin{align}\nonumber
&\prod_{i,j\geq 1}
\frac{f(\mu_{1,i}-\mu_{2,j}+j-i)}{f(j-i)}\\
=&\prod_{i=1}^{\len{\R_1}}\prod_{j=1}^{\len{\R_2}}
\frac{f(\mu_{1,i}-\mu_{2,j}+j-i)}{f(j-i)}
\prod_{i=1}^{\len{\R_1}}
\prod_{v=1}^{\mu_{1,i}}
\frac{1}{f(v-i+\len{\R_1})}
\prod_{j=1}^{\len{\R_2}}
\prod_{v=1}^{\mu_{2,j}}
\frac{1}{f(-v+j-\len{\R_2})}
\\
=&\prod_{i,j\geq 1}
\frac{f(\mu_{1,i}+\mu_{2,j}^{\vee}-i-j+1)}{f(-i-j+1)}
\label{end}
\\\label{nak}
=&\prod_{(i,j)\in\R_1}\frac{1}
{f({\mu_{1,i}-j+\mu_{2,j}^{\vee}-i+1})}
\prod_{(i,j)\in\R_2}\frac{1}
{f({-\mu_{1,j}^{\vee}+i-\mu_{2,i}+j-1})}
.
\end{align}
The  proof of  the last expression (\ref{nak})
is the same as the proof of 
theorem 1.11 in \cite{NaYo}.
The proofs of 
other formulas (\ref{start})-(\ref{end})
can be found in \cite{Zhou1}.

\section{Calculation of the 
integer $\boldsymbol{m_i}$}
\begin{figure}[th]
\begin{center}
\unitlength 0.1in
\begin{picture}( 29.0000, 26.3000)( 14.4000,-32.0000)
%
\special{pn 8}%
\special{pa 1600 2000}%
\special{pa 2800 2000}%
\special{fp}%
\special{sh 1}%
\special{pa 2800 2000}%
\special{pa 2734 1980}%
\special{pa 2748 2000}%
\special{pa 2734 2020}%
\special{pa 2800 2000}%
\special{fp}%
%
\special{pn 8}%
\special{pa 1600 2000}%
\special{pa 2000 800}%
\special{fp}%
\special{sh 1}%
\special{pa 2000 800}%
\special{pa 1960 858}%
\special{pa 1984 852}%
\special{pa 1998 870}%
\special{pa 2000 800}%
\special{fp}%
%
\special{pn 8}%
\special{pa 1600 2000}%
\special{pa 2600 3200}%
\special{fp}%
\special{sh 1}%
\special{pa 2600 3200}%
\special{pa 2574 3136}%
\special{pa 2566 3160}%
\special{pa 2542 3162}%
\special{pa 2600 3200}%
\special{fp}%
%
\special{pn 8}%
\special{pa 2000 800}%
\special{pa 2800 2000}%
\special{fp}%
%
\special{pn 8}%
\special{pa 2810 2000}%
\special{pa 2600 3200}%
\special{fp}%
%
\special{pn 8}%
\special{pa 2200 1600}%
\special{pa 2200 2400}%
\special{fp}%
%
\special{pn 8}%
\special{pa 2200 1600}%
\special{pa 2800 1200}%
\special{fp}%
%
\special{pn 8}%
\special{pa 2200 2400}%
\special{pa 3400 2600}%
\special{fp}%
%
\special{pn 8}%
\special{pa 2200 2400}%
\special{pa 1700 3000}%
\special{fp}%
%
\special{pn 8}%
\special{pa 1600 1400}%
\special{pa 2200 1600}%
\special{fp}%
\put(26.4000,-33.6000){\makebox(0,0)[lb]{$v_{i+1}$}}%
\put(28.7000,-21.0000){\makebox(0,0)[lb]{$v_i$}}%
\put(19.3000,-7.4000){\makebox(0,0)[lb]{$v_{i-1}$}}%
%
\special{pn 13}%
\special{pa 2200 2400}%
\special{pa 2680 2480}%
\special{fp}%
\special{sh 1}%
\special{pa 2680 2480}%
\special{pa 2618 2450}%
\special{pa 2628 2472}%
\special{pa 2612 2490}%
\special{pa 2680 2480}%
\special{fp}%
\put(22.9000,-26.6000){\makebox(0,0)[lb]{$v_{\rm in}$}}%
%
\special{pn 13}%
\special{pa 2200 1600}%
\special{pa 1720 1440}%
\special{fp}%
\special{sh 1}%
\special{pa 1720 1440}%
\special{pa 1778 1480}%
\special{pa 1772 1458}%
\special{pa 1790 1442}%
\special{pa 1720 1440}%
\special{fp}%
\put(19.3000,-14.7000){\makebox(0,0)[lb]{$v_{\rm out}$}}%
\put(14.4000,-21.0000){\makebox(0,0)[lb]{$\vec{0}$}}%
%
\special{pn 8}%
\special{pa 2740 2490}%
\special{pa 2750 2490}%
\special{pa 2750 2370}%
\special{pa 2740 2370}%
\special{pa 2740 2490}%
\special{fp}%
\put(36.9000,-12.8000){\makebox(0,0)[lb]{$-\gamma_i v_i=v_{i+1}+v_{i-1}$}}%
%
\special{pn 8}%
\special{ar 2850 2310 1530 550  4.2872674 5.9965926}%
%
\special{pn 8}%
\special{pa 2240 1806}%
\special{pa 2220 1810}%
\special{fp}%
\special{sh 1}%
\special{pa 2220 1810}%
\special{pa 2288 1820}%
\special{pa 2272 1802}%
\special{pa 2282 1780}%
\special{pa 2220 1810}%
\special{fp}%
\put(43.4000,-22.9000){\makebox(0,0)[lb]{$m_i=-\gamma_i-1$}}%
%
\special{pn 8}%
\special{pa 2200 1680}%
\special{pa 2200 2240}%
\special{fp}%
\special{sh 1}%
\special{pa 2200 2240}%
\special{pa 2220 2174}%
\special{pa 2200 2188}%
\special{pa 2180 2174}%
\special{pa 2200 2240}%
\special{fp}%
\end{picture}%
\end{center}
\caption{$m_i=\det(v_{\rm in},v_{\rm out})=-\gamma_i-1.$}
\label{picm}
\end{figure}
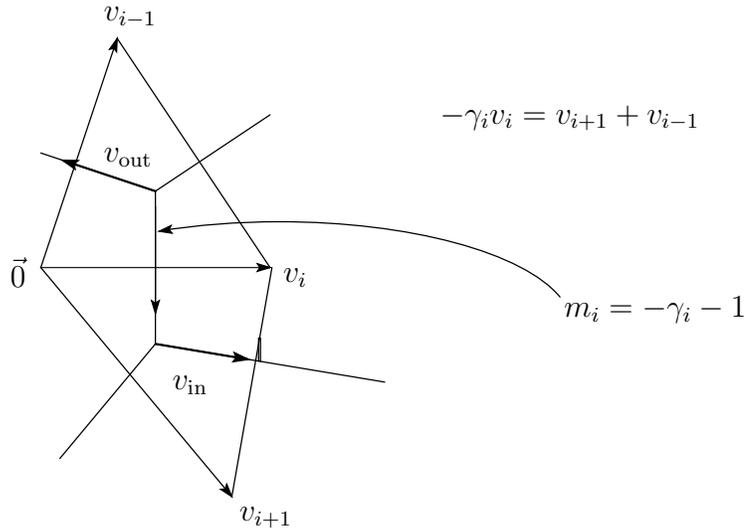

\hspace*{2ex}
In this section, we present the calculation of (\ref{m}).

By definition, $m_i=\det(v_{\rm in},v_{\rm out})$
where $v_{\rm in}$ and $v_{\rm out}$
are two-component vectors 
described in figure \ref{picm}.
Note that 
$v_{\rm out}\perp v_{i-1}$, 
$v_{\rm in}\perp (v_{i+1}-v_{i})$
and $\det(v_{i-1},v_{\rm out})=
\det(v_{i+1}-v_i,v_{\rm in})=1$.
Therefore
\begin{equation}
\begin{split}
\det(v_{\rm in},v_{\rm out})&=\det
(v_{i+1}-v_i,v_{i-1})
=\det(v_{i+1},v_{i-1})-\det(v_{i},v_{i-1}).
\end{split}
\end{equation}
The second term is minus the volume of the triangle
spanned by $v_i,v_{i-1}$, which is $-1$ because the
surface is smooth.
To compute the first term, note that 
the following equation holds by (\ref{SI}):
\begin{equation*}
-\gamma_i\det(v_{i+1},v_{i})=\det(v_{i+1},v_{i-1}).
\end{equation*}
Since $\det(v_{i+1},v_i)=1$, the right-hand side is 
equal to $-\gamma_i$.
Therefore
\begin{equation*}
m_i=\det(v_{\rm in},v_{\rm out})=-\gamma_i-1.
\end{equation*}



\end{document}